\documentclass[5p,twocolumn,times,number]{elsarticle}

\usepackage{graphicx}
\usepackage{amsmath}
\usepackage{lineno}
\usepackage[compact]{titlesec}

\def\pmbanner{{\hrule height 1 pt}\vskip25pt{PUBLISHED IN NIMA (2019)}\vskip25pt{\hrule height 4pt}\vskip15pt}
\begin{document}

\begin{frontmatter}

%% Note: \pmbanner before the actual title
\title{\pmbanner The new drift chamber of the {MEG} II experiment}

\author[infnpisa,unisi]{M.~Chiappini\corref{cor}}
\ead{marco.chiappini@pi.infn.it}
\author[infnpisa]{A.~M.~Baldini}
\author[infnroma,sapienza]{G.~Cavoto}
\author[infnpisa,unipi]{F.~Cei}
\author[infnroma,sapienza]{G.~Chiarello}
\author[infnpisa,unipi]{M.~Francesconi}
\author[infnpisa]{L.~Galli}
\author[infnlecce]{F.~Grancagnolo}
\author[infnpisa]{M.~Grassi}
\author[psi]{M.~Hildebrandt}
\author[infnpisa,unipi]{D.~Nicol\`o}
\author[infnlecce,unile]{M.~Panareo}
\author[infnpisa,unipi,psi]{A.~Papa}
\author[infnpisa]{F.~Raffaelli}
\author[infnroma,frascati]{F.~Renga}
\author[infnpisa]{G.~Signorelli}
\author[infnlecce,unile]{G.~F.~Tassielli}
\author[infnroma]{C.~Voena}

\cortext[cor]{Corresponding author}

\address[infnpisa]{INFN Sezione di Pisa, Largo B. Pontecorvo 3, 56127, Pisa, Italy}
\address[unisi]{Dipartimento di Scienze Fisiche, della Terra e dell'Ambiente dell'Universit\`a, Via Roma 56, 53100, Siena, Italy}
\address[infnroma]{INFN Sezione di Roma, Piazzale A. Moro 2, 00185, Roma, Italy}
\address[sapienza]{Dipartimento di Fisica dell'Universit\`a ``Sapienza'' di Roma, Piazzale A. Moro 2, 00185, Roma, Italy}
\address[unipi]{Dipartimento di Fisica dell'Universit\`a di Pisa, Largo B. Pontecorvo 3, 56127, Pisa, Italy}
\address[infnlecce]{INFN Sezione di Lecce, Via per Arnesano, 73100, Lecce, Italy}
\address[psi]{Paul Scherrer Institut PSI, 5232, Villigen, Switzerland}
\address[unile]{Dipartimento di Matematica e Fisica dell'Universit\`a del Salento, Via per Arnesano, 73100, Lecce, Italy}
\address[frascati]{Laboratori Nazionali di Frascati, Via Enrico Fermi 40, 00044, Frascati (Rome), Italy}

\begin{abstract}

%was built by INFN Pisa, Lecce and Rome groups and
%Moreover, minimization of background $\gamma$ rate in the LXe calorimeter from bremsstrahlung and $\mbox{e}^+$ Annihilation-In-Flight (AIF) with $\mbox{e}^-$ of traversed materials is of crucial importance.

This article presents the MEG II Cylindrical Drift CHamber (CDCH), a key detector for the phase 2 of MEG, which aims at reaching a sensitivity level of the order of $6 \times 10^{-14}$ for the charged Lepton Flavour Violating $\mu^+ \rightarrow \mbox{e}^+ \gamma$ decay \cite{nuovocimento} \cite{cei}. CDCH is designed to overcome the limitations of the MEG $\mbox{e}^+$ tracker \cite{megup} and guarantee the proper operation at high rates with long-term detector stability. CDCH is a low-mass unique volume detector with high granularity: 9 layers of 192 drift cells, few mm wide, defined by $\approx 12000$ wires in a stereo configuration for longitudinal hit localization. The total radiation length is $1.5 \times 10^{-3}$ $\mbox{X}_0$, thus minimizing the Multiple Coulomb Scattering (MCS) contribution and allowing for a single-hit resolution of 110 $\mu$m \cite{singlehit} and a momentum resolution of 130 keV/c. CDCH integration into the MEG II experimental apparatus \cite{meg2} will start in this year.

%CDCH is currently in the final mechanical and HV test phase before the shipping to Paul Scherrer Institut (PSI) for commissioning and integration into the MEG II experimental apparatus \cite{meg2}.

\end{abstract}


\end{frontmatter}

\section{Introduction}

%and is presently under construction

The MEG experiment represents the state of the art in the search for the charged Lepton Flavour Violating $\mu^+ \rightarrow \mbox{e}^+ \gamma$ decay \cite{nuovocimento} \cite{cei}. The MEG collaboration presented the final results for the phase 1 of the experiment \cite{meg1} exploiting the full statistics collected during the 2009-2013 data taking period at Paul Scherrer Institut (PSI). The final analysis resulted in the new world best upper limit on the BR$(\mu^+ \rightarrow \mbox{e}^+ \gamma) < 4.2 \times 10^{-13}$ (90\% C.L.). The MEG experiment has reached its ultimate level of sensitivity, limited by the resolutions on the measurement of the kinematic variables of the two decay products. Therefore an upgrade of the experimental apparatus (MEG II) \cite{megup} has been approved. MEG II aims at reaching a sensitivity enhancement of one order of magnitude compared with the final MEG results, in 3 years of data taking, by improving the detector figures of merit and increasing the $\mu^+$ stopping rate, up to $7 \times 10^{7}$ $\mu^+$/s (double than MEG).

\subsection{The MEG II $\mbox{e}^+$ spectrometer}

The MEG II detector was optimized to fulfill the fundamental requirements of high transparency for 50 MeV $\mbox{e}^+$, fast response and stable operation at high rates for precision measurements and background rejection \cite{meg2}. The MEG II $\mbox{e}^+$ spectrometer consists in a low-mass Cylindrical Drift CHamber (CDCH) with high granularity and stereo wires configuration, followed by a pixelated Timing Counter (pTC), based on scintillator tiles read out by SiPMs, for precise measurement of $\mbox{e}^+$ momentum vector and time respectively. Both detectors are placed inside the COBRA superconducting gradient field magnet.

%designed specifically for the MEG experiment.
%In the new configuration the amount of material crossed by the particles along their helix trajectories is extremely reduced, allowing a more accurate determination of the decay $\mbox{e}^+$ kinematic variables.

\section{The Cylindrical Drift CHamber (CDCH)}

%extending radially from 17 cm to 29 cm and for 1.91 m in length
%The wires choice is dictated by keeping MCS at minimum.
%($\mbox{X}_0 \sim 5300$ m)
%lowers the mixture $\mbox{X}_0$ to 1300 m, but
%The use of He allows to minimize MCS.
%In this way
%$2\pi$
%whose surface is given by the envelope of the wires planes
%azimuthally divided in 12 30$^{\circ}$ sectors, 16 drift cells wide

\begin{figure}[h]
\centering
\includegraphics[width=.99\linewidth]{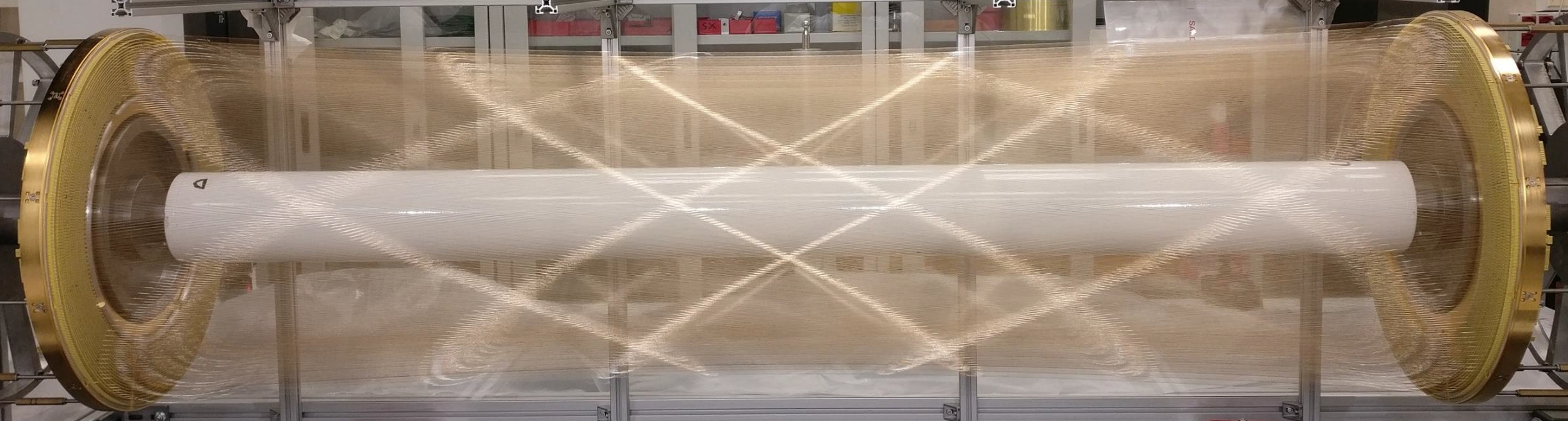}
\caption{The fully wired MEG II CDCH. The hyperbolic side profile is visible.}
\label{fig:cdch}
\end{figure}

CDCH is a single volume detector with a cylindrical symmetry along the $\mu^+$ beam. The length is 1.91 m and the radial width ranges from 17 to 29 cm. The full azimuthal coverage around the $\mu^+$ stopping target is guaranteed. This improves the geometric acceptance for signal $\mbox{e}^+$ and allows to use new tracking procedures capable to exploit $\times 4$ hits more than MEG and match the information reconstructed by CDCH and pTC for a larger tracking efficiency ($\approx 80$\%). The high granularity is ensured by 9 layers of 192 drift cells. Each layer consists of 2 criss-crossing field wires planes enclosing a sense wires plane. The wires are not parallel to CDCH axis, but form an angle varying from 6$^{\circ}$ in the innermost layer to 8.5$^{\circ}$ in the outermost one. The stereo angle has an alternating sign, depending on layer, allowing to reconstruct the longitudinal hit coordinate. Since the wires will be readout at both ends, this measurement will be improved by using the techniques of charge division and time propagation difference. The stereo configuration gives CDCH the shape of a rotation hyperboloid. The single drift cell is quasi-square with a 20 $\mu$m Au-plated W sense wire surrounded by 40/50 $\mu$m Ag-plated Al field wires, with 5:1 field-to-sense wires ratio and a total number of $\approx 12000$ wires (Figure \ref{fig:cdch}).

The cell width increases linearly with the radius and also slightly varies along CDCH axis. Near CDCH endplates the width ranges from 6.7 to 8.7 mm, while at the center from 5.8 to 7.5 mm. The sensitive volume is filled with a low-mass He:i$\mbox{C}_4\mbox{H}_{10}$ (85:15) gas mixture \cite{gas}. This gas choice is the best compromise between high transparency and single-hit resolution \cite{singlehit}. The addition of a small amount of isobutane is required as a quencer to avoid self-sustained discharges. The number of $\mbox{e}^-$-ion pairs per crossing particle is $4.8 \mbox{ cm}^{-1}$. All the aforementioned features contribute to minimize the amount of material crossed by $\mbox{e}^+$ along their helix trajectories, thus reducing the MCS contribution and the total radiation length down to $1.5 \times 10^{-3}$ $\mbox{X}_0$ per track turn (in MEG it was $2 \times 10^{-3}$ $\mbox{X}_0$). A single-hit resolution of 110 $\mu$m was measured in a dedicated facility on prototypes \cite{singlehit}. Several MC studies show a momentum resolution in agreement with MEG II experimental requirements: 130 keV/c with full signal + background events. In MEG II a particle flux double than MEG is foreseen and CDCH must withstand it. The occupancy is maximum in the core, at inner radii near the stopping target ($\mbox{max} \approx 23$ kHz/cm) and decreases while radius and longitudinal coordinate increase ($\mbox{min} < 5$ kHz/cm). Ageing tests with prototypes were performed, resulting in a gain loss $< 10$\% per DAQ year for the inner layers. A fast front-end electronics with a total bandwidth of nearly 1 GHz will be used \cite{meg2}.

%This small number, together with small drift cells, lead to a bias in the estimate of the $\mbox{e}^+$ impact parameter coming from the statistical fluctuation of the primary ionization clusters along the track. The use of fast front-end electronics, with a total bandwidth of nearly 1 GHz, capable of timing all arriving ionisation clusters will allow to obtain significant improvements.

\section{CDCH construction and status}

%in proper slots of CDCH endplates

\begin{figure}[h]
\centering
\includegraphics[width=.502\linewidth]{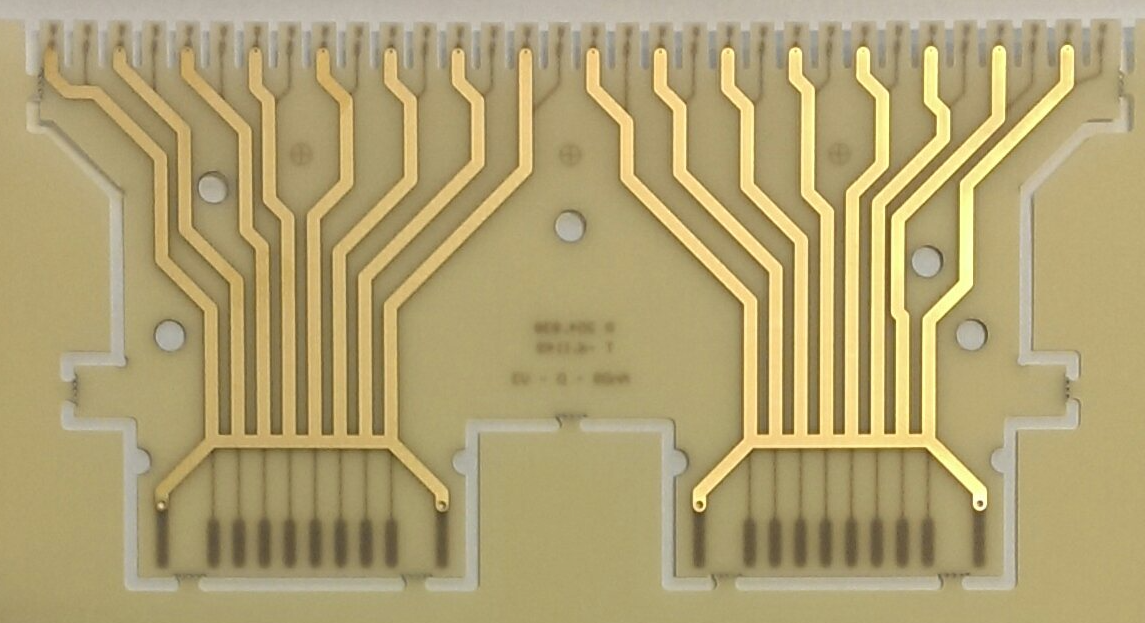}
\includegraphics[width=.364\linewidth]{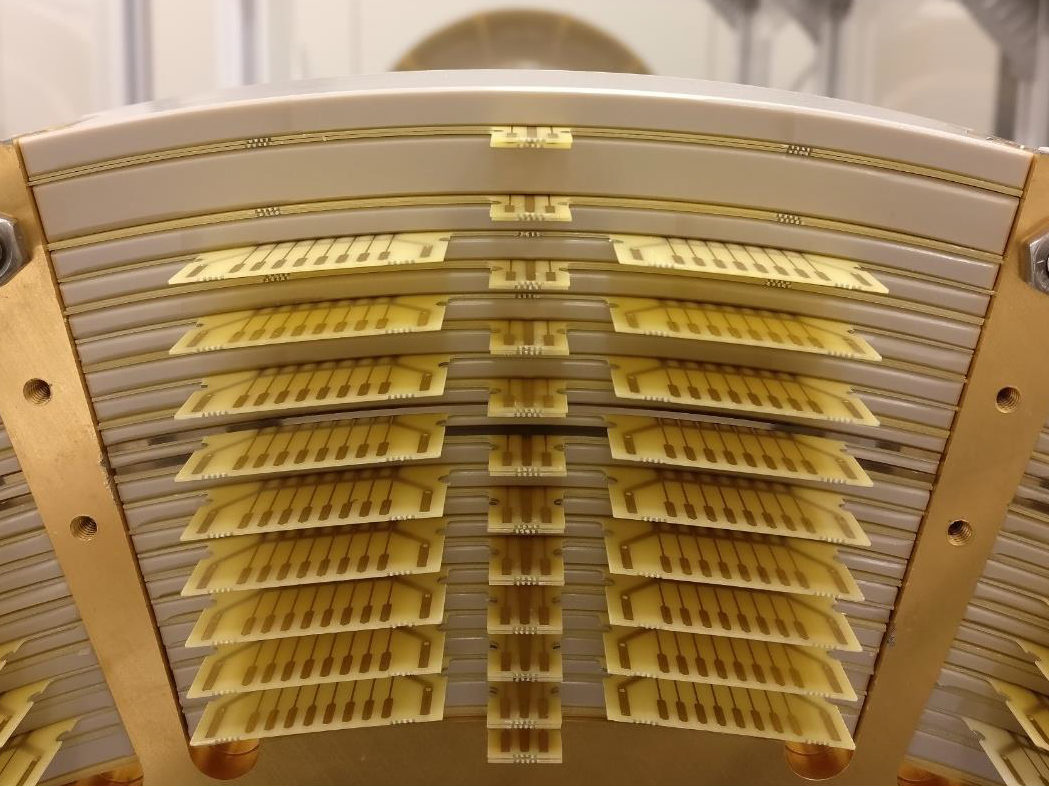}
\caption{Left: one of the PCBs where wires are soldered. Right: wire-PCBs stack with PEEK spacers between the spokes of CDCH endplate.}
\label{fig:pcbs}
\end{figure}

CDCH design and construction involved the common effort of 3 italian working groups within the MEG collaboration: INFN of Pisa, Lecce and Rome. CDCH is the first drift chamber ever designed and built in a modular way. In fact, given the high wires density (12 wires/$\mbox{cm}^2$), the classical technique with wires anchored to endplates with feedthroughs is hard to implement. The wires are not strung directly on the final chamber, but they are soldered at both ends on the pads of two PCBs, which are then mounted on CDCH. The wiring procedure is done in Lecce with an automatic robot which fixes the wires on PCBs with a contactless laser soldering. CDCH is then assembled in Pisa by radially overlapping the wire-PCBs in the 12 30$^{\circ}$ sectors of the helm-shaped endplates, between the spokes which act as housing for the wire-PCBs \cite{meg2}. Each wire-PCB is placed at the proper radius through PEEK spacers whose thickness is adjusted to have the correct radial dimension of the drift cells (Figure \ref{fig:pcbs}). All the operations are performed inside cleanrooms.

\begin{figure}[h]
\centering
\includegraphics[width=.382\linewidth]{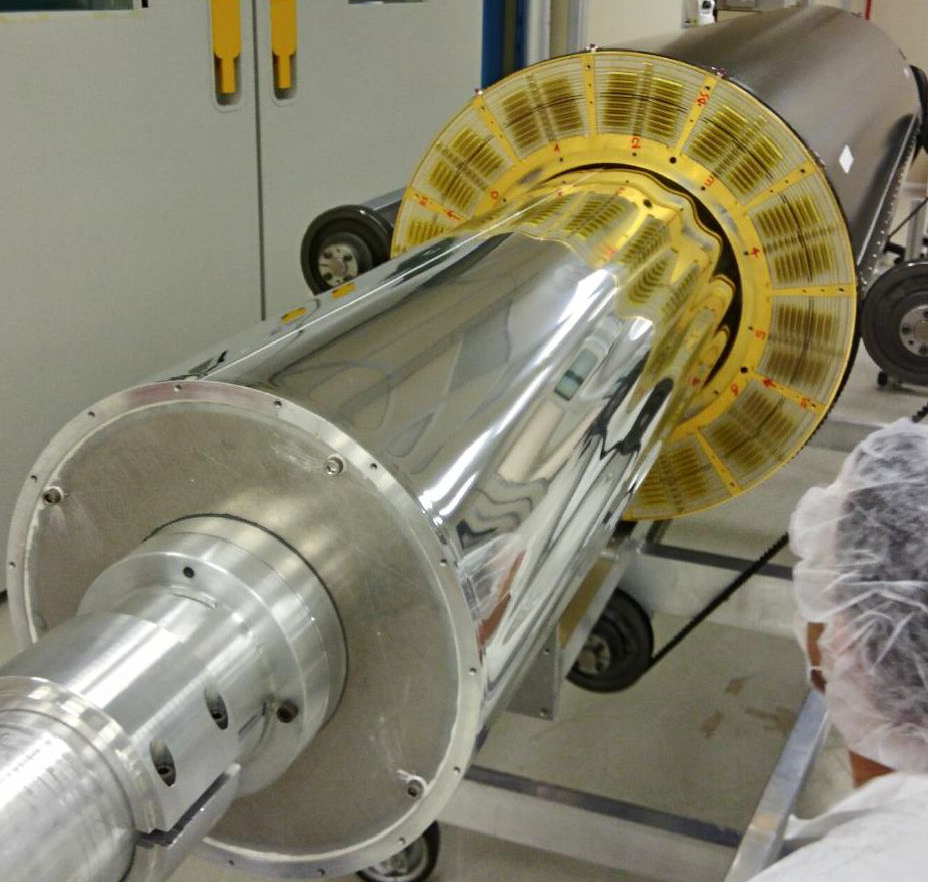}
\includegraphics[width=.30\linewidth]{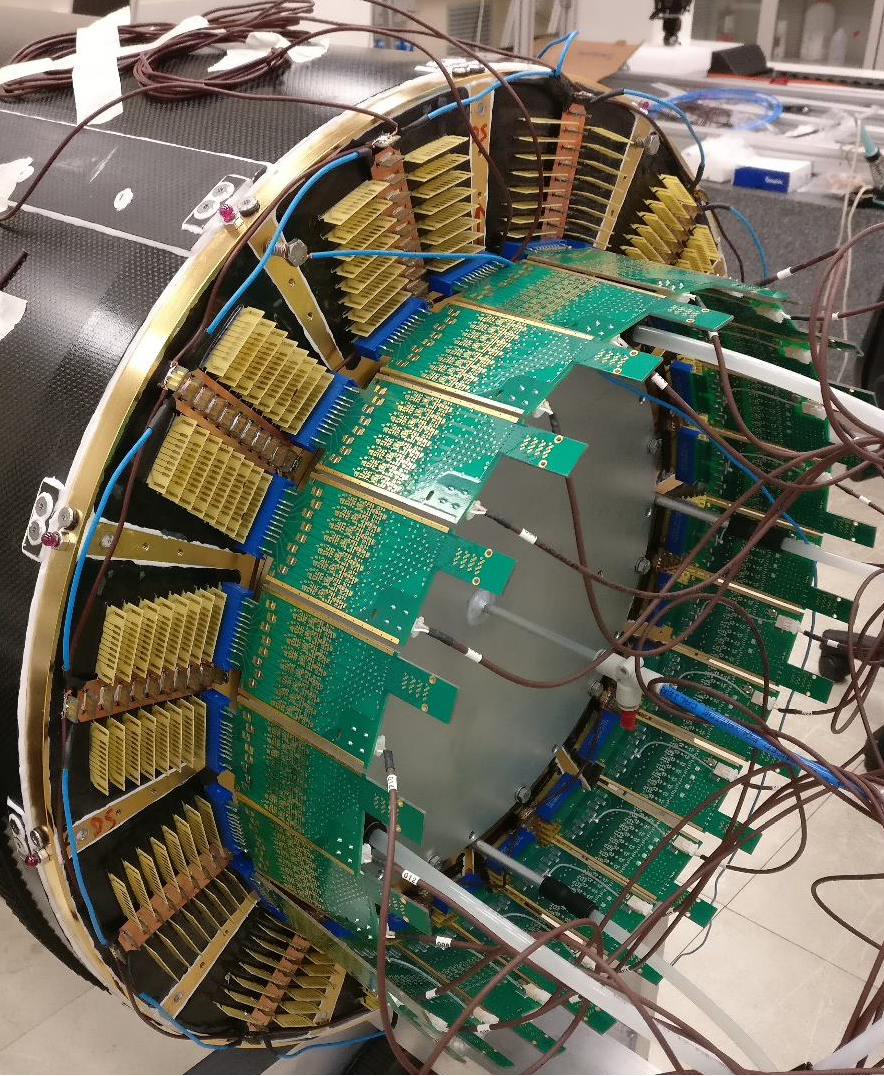}
\caption{Left: Insertion phase of the 20 $\mu$m Al mylar foil. On CDCH side the carbon fiber support structure is visible. Right: One phase of the HV test.}
\label{fig:closing&test}
\end{figure}

At the innermost radius a 20 $\mu$m one-side-Al mylar foil separates CDCH gas volume from the He filled target region. At the outermost radius a carbon fiber support structure encloses the sensitive volume and keeps the endplates at the correct distance, ensuring the proper mechanical wires tension (Figure \ref{fig:closing&test}).

CDCH is currently in the final mechanical and HV test phase before the shipping to PSI for commissioning and integration into the MEG II experimental apparatus \cite{meg2}.

%% bibliography


\begin{thebibliography}{10}
\expandafter\ifx\csname url\endcsname\relax
  \def\url#1{\texttt{#1}}\fi
\expandafter\ifx\csname urlprefix\endcsname\relax\def\urlprefix{URL }\fi

\bibitem{nuovocimento}
L.~Calibbi and G.~Signorelli, Charged lepton flavour violation: An experimental and theoretical introduction,
  La Rivista del Nuovo Cimento Volume 41 Serie 5 Numero 2 (2018).
  
\bibitem{cei}
F.~Cei and D.~Nicol\`o, Lepton Flavour Violation Experiments, Advances in High Energy Physics, Article ID 282915 (2014).
  
\bibitem{meg1}
A.~M.~Baldini, et~al., Search for the lepton flavour violating decay $\mu^+ \rightarrow e^+ \gamma$ with the full dataset of the {MEG} experiment,
  Eur. Phys. J. C (2016) 76--434.
  
\bibitem{megup}
A.~M.~Baldini, et~al., {MEG} Upgrade Proposal,
  arXiv:1301.7225. %[physics.ins-det].
  
\bibitem{meg2}
A.~M.~Baldini, et~al., The design of the {MEG} II experiment,
  Eur. Phys. J. C (2018) 78--380.
  
\bibitem{singlehit}
A.~M.~Baldini, et~al., Single-hit resolution measurement with {MEG} II drift chamber prototypes,
  JINST 11 P07011 (2016).
  
\bibitem{gas}
A.~M.~Baldini, et~al., Gas distribution and monitoring for the drift chamber of the MEG II experiment,
  JINST 13 P06018 (2018).
  
%\bibitem{clustertime}
%M.~Cascella, F.~Grancagnolo, G.~Tassielli, Cluster Counting/Timing Techniques for Drift Chambers,
%  Nuclear Physics B (Proc. Suppl.) 248--250 (2014) 127--130.


\end{thebibliography}
\end{document}